\documentclass[3p, authoryear]{elsarticle}

\usepackage[T1]{fontenc}
\usepackage[utf8x]{inputenc}

\usepackage{microtype}

\usepackage{graphicx}
\usepackage[export]{adjustbox}[2011/08/13]
\usepackage[labelfont=bf]{caption}
\usepackage{floatrow}
\usepackage{amsmath}
\usepackage{siunitx}
\usepackage{booktabs}
\usepackage[normalem]{ulem}

\usepackage{pdfpages}

\newcommand{\mytitle}{A parameter-free mechanistic model of the adhesive wear process of rough surfaces in sliding contact}

\usepackage{hyperref} %
\hypersetup{
  final,
  pdfborder={0 0 0},
  pdftitle={\mytitle},
  pdfauthor={Brink, Frérot, Molinari},
  colorlinks=true,
  urlcolor=blue,
  linkcolor=blue,
  citecolor=blue
}

\begin{document}
\frenchspacing

\begin{frontmatter}
\title{\mytitle}

\author{Tobias Brink\corref{cor1}\fnref{mpie}}
\ead{t.brink@mpie.de}

\author{Lucas Frérot\corref{corXX}\fnref{jhu}}

\author{Jean-François Molinari\corref{cor2}}
\ead{jean-francois.molinari@epfl.ch}

\cortext[cor1]{Corresponding author}
\address{Civil Engineering Institute and Institute of Materials
  Science and Engineering,\\\'Ecole polytechnique f\'ed\'erale de
  Lausanne (EPFL), Station 18, CH-1015 Lausanne, Switzerland}
\fntext[mpie]{Present address: \textit{Max-Planck-Institut f\"ur
  Eisenforschung GmbH, Max-Planck-Stra\ss{}e 1, 40237 D\"usseldorf,
  Germany}}
\fntext[jhu]{Present address: \textit{Dept. of Physics and Astronomy,
    Bloomberg Center for Physics and Astronomy, Johns Hopkins
    University, 3400 Charles St., Baltimore, MD 21218, USA}}

\journal{J.~Mech.~Phys.~Solids, online at
  \href{https://doi.org/10.1016/j.jmps.2020.104238}
       {https:/\!/doi.org/10.1016/j.jmps.2020.104238}.}

\date{\today}
\begin{abstract}
  In order to develop predictive wear laws, relevant material
  parameters and their influence on the wear rate need to be
  identified. Despite decades of research, there is no agreement on
  the mathematical form of wear equations and even the simplest
  models, such as Archard's, contain unpredictable fit
  parameters. Here, we propose a simple model for adhesive wear in dry
  sliding conditions that contains no fit parameters and is only based
  on material properties and surface parameters. The model connects
  elastoplastic contact solutions with the insight that volume
  detachment from sliding surfaces occurs in the form of wear
  particles, the minimum size of which can be estimated. A novelty of
  the model is the explicit tracking of the sliding process, which
  allows us to meaningfully connect particle emission rates and sizes
  to the macroscopic wear rate. The results are qualitatively
  promising, but we identify the necessity for more controlled wear
  experiments and the parameters needed from such work in order to
  fully verify and improve our model.
\end{abstract}


\end{frontmatter}

\vspace{-\baselineskip} %

\section{Introduction}

In engineering practice, wear is usually predicted by (semi-)empirical
laws instead of mechanistically informed models. Unfortunately, these
laws are numerous and not necessarily transferable, while experiments
which reveal sufficient information to validate them are scarce
\citep{Meng1995}. In order to further our fundamental understanding of
the wear process, parameter-free wear predictions based on microscopic
mechanisms are needed. The main obstacle in this path is the multitude
of those mechanisms~\citep[reviewed in][]{Vakis2018}, spanning from
corrosion, local melting, subsurface plasticity, and phase
transformation to fatigue and fracture from the nanoscale to the
macroscale. A universal, multi-physics description of this process is
thus currently out of reach and scientific progress in this area
relies on describing idealized wear modes.

Hence, our focus lies here on the case of adhesive wear of brittle
materials in the unlubricated regime. Adhesive wear is perhaps one of
the most common types of wear, in which wear particles detach at
points where strong enough junctions form between asperities from two
contacting surfaces sliding relative to one another
\citep{Rabinowicz1995}. The earliest works on wear already recognized
that the wear volume $V$---i.e., the loss of volume of relatively
sliding surfaces---is proportional to the applied normal force $F_N$
and the relative sliding distance $s$, at least in a certain load
range \citep{Reye1860, Rabinowicz1951}. The value of interest is
therefore the prediction of the wear rate
$W = \mathrm{d}V/\mathrm{d}s$. \Citet{Bowden1939} provided the
critical insight that all surfaces are rough at some level and that
the value of the real contact area thus depends on $F_N$. With this,
\citet{Holm1967}, \citet{Burwell1952}, and \citet{Archard1953}
developed wear equations of the form
\begin{equation}
    \label{eq:Archard-law}
    W = k\dfrac{F_N}{\tilde{p}},
\end{equation}
which assume that the wear rate is proportional to the real contact
area given by $F_N$ divided by the material's hardness $\tilde{p}$
with some proportionality constant $k$. The contribution of
\citet{Archard1953} is the insight that wear progresses due to the
detachment of wear particles at contact junctions between the surfaces
instead of by atom-by-atom attrition. Nevertheless, the predictive
power of Eq.~\ref{eq:Archard-law} is low, since the constant $k$,
commonly called ``wear coefficient'', remains an empirical parameter
whose values can span several orders of magnitude.

While \citet{Archard1953} treated the wear particle formation as a
random event with $k$ as the detachment probability, recent computer
modeling efforts \citep{Aghababaei2016, Aghababaei2017,
  Aghababaei2019a, Brink2019} confirmed and expanded on an old
hypothesis by \citet{Rabinowicz1958} that there is a critical length
scale for wear particle formation. If contact junctions are smaller
than this critical size, they do not detach from the surface, but show
some sort of plastic activity. The model is based on the amount of
energy that can be stored locally in a contact and is thus available
to propagate the crack that leads to particle detachment. The critical
length scale relates to a value for the junction diameter $d$ and can
be expressed, based on the material's properties, as
\citep{Aghababaei2016}
\begin{equation}
  \label{eq:d-star}
  d^* = f \dfrac{\mathcal{G}}{\tilde{\tau}^2 / 2\mu},
\end{equation}
where $f$ is a geometry factor (often set to $\approx 3$, see
\citet{Aghababaei2016} and \citet{Brink2019} for a
detailed discussion and derivation), $\mathcal{G}$ is the fracture
energy, and $\tilde{\tau}^2 / 2\mu$ is the elastic energy density of an
asperity with shear modulus $\mu$ when it is loaded to its
shear limit $\tilde{\tau}$. It should be noted that this concept has
only been tested with relatively brittle material models and an
extension to materials with ductile fracture behavior is outstanding,
hence our limitation to the brittle regime in the present paper.

A first approach to advance Archard's model based on this
deterministic particle formation criterion was developed by
\citet{Frerot2018} using elastic contact solutions of synthesized,
self-affine, rough surfaces. There, $k$ is no longer a detachment
probability, but related to the fraction of asperities with
$d \geq d^*$. This model is revisited in section~\ref{sec:repeat-Frerot},
so we simply summarize that it does not reproduce a wear coefficient
that is constant over a large load range as experimentally observed
\citep[e.g., by][]{Archard1956}. Another approach by
\citet{Popov2018a} is also based on elastic contact solutions of
synthesized, self-affine, rough surfaces, but tries to avoid defining
an asperity or contact junction directly, instead applying the
energetic criterion underlying Eq.~\ref{eq:d-star} to arbitrary areas
of the surface. This model predicts that infinitely large wear
particles can form and yields superlinear wear rates even when
imposing an upper limit on particle size. The reason is (most likely)
that having sufficient elastic energy available is only a necessary
condition for fracture and wear particle detachment; as already
implicit in the fracture model of \citet{Griffith1921}, there must
also always be a stress concentration to propagate a crack.

Based on these results, it is not yet clear if the critical length
scale of Eq.~\ref{eq:d-star} is helpful in predicting macroscopic wear
rates. Indeed, the value of $d^*$ decreases with increasing hardness
of the material, which might suggest that a harder material can more
easily produce wear debris even from small contacts and thus wear
more, in contradiction to the typical experimental observation. The
mechanistic model by \citet{Frerot2018} shares this shortcoming: Wear
rates increase with decreasing $d^*$.

The goal of the present contribution is therefore to show that a
critical ingredient of the wear process was not considered in
sufficient detail before: The surfaces slide relative to each other
during wear. The explicit inclusion of a very simple facsimile of the
sliding process into our model allows us to make sense of the seeming
contradiction between the microscopic length scale $d^*$ and the
macroscopic wear laws. Furthermore, we find that the material
properties used to calculate $d^*$ cannot be divorced from the physics
of the contact solution and that the relation between wear particle
formation and wear rate is nontrivial. A comparison with an
experimental work by \citet{Kim1986} that contains sufficient data to
feed our model yields some qualitative agreement, but ultimately more
detailed experiments are needed to develop predictive wear models.

\section{Complementing a static model with plasticity}
\label{sec:repeat-Frerot}

\begin{figure}
  \centering
  \includegraphics[center]
                  {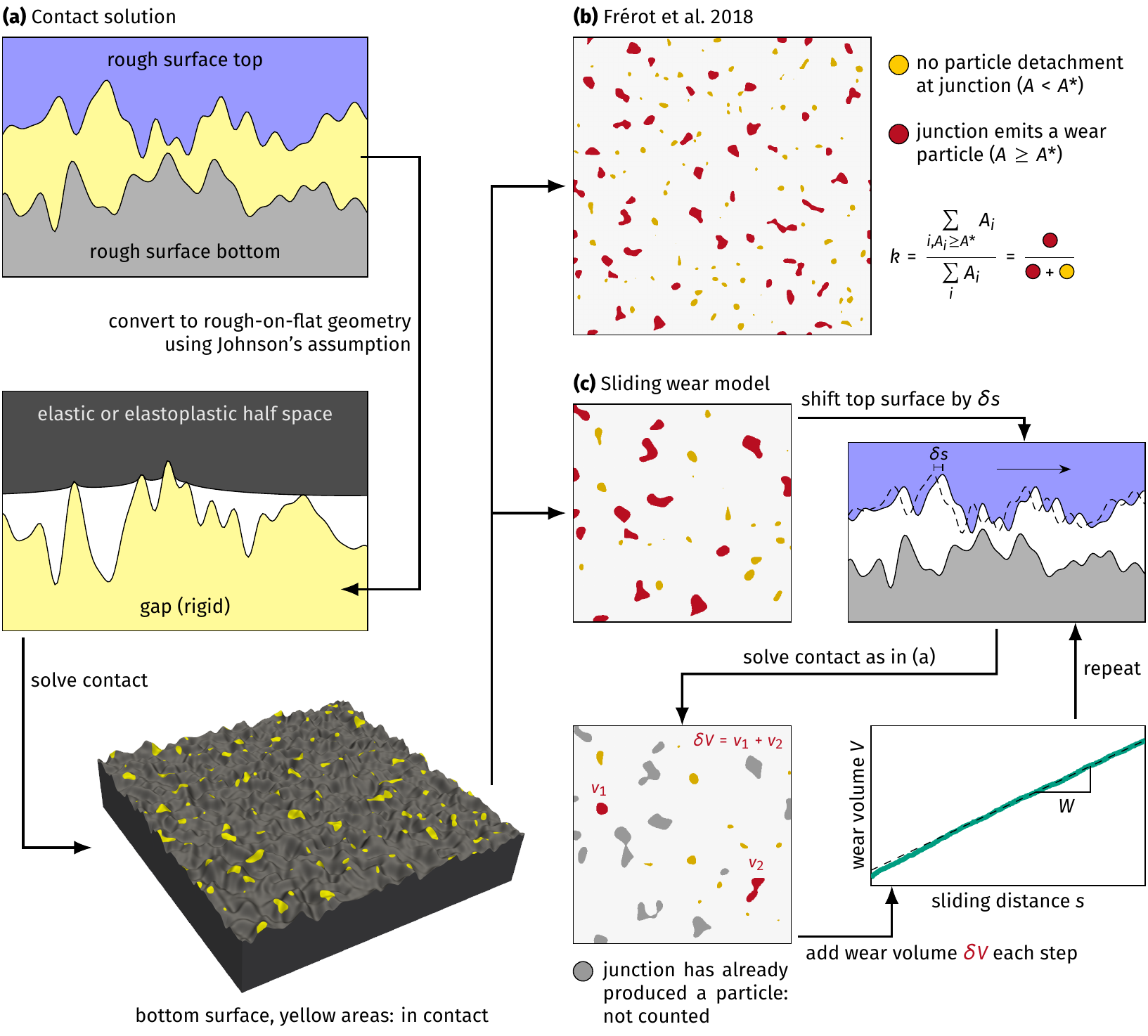}
  \caption{Schematic of the wear models. (a) The models in this work
    are based on the elastoplastic contact solution of two rough
    surfaces, which is approximated by solving the problem of the
    contact between a rigid surface (whose geometry is equal to the
    gap between the two original surfaces) and a flat
    elastic-perfectly plastic half space~\citep{Johnson1985}.  From
    this, we obtain the contact junctions shown in yellow at the
    bottom. (b) A first model proposed by \citet{Frerot2018} states
    that Archard's wear coefficient $k$ is equal to the ratio of total
    area of the junctions large enough to form wear particles divided
    by the real contact area. (c) Our improved model takes the
    evolution of contacts during the sliding process into account. The
    top surface is shifted by a small length $\delta s$ each
    step. Similar to \citet{Frerot2018}, we mark junctions that form
    wear particles. In contrast to the earlier work, we ignore
    junctions that have formed a wear particle in a previous step. All
    others contribute a total wear volume $\delta V$ each step. After
    a running-in period, the wear rate can be estimated as
    $W = \mathrm{d} V / \mathrm{d} s$. Note that the $\delta s$
    between the two contact maps shown here is quite large in order to
    make the evolution more easily visible.}
  \label{fig:illustrate-method}
\end{figure}

Our proposed new wear modeling approach is based on the work of
\citet{Frerot2018}, the basic ingredient of which is an elastic
contact solution. Figure~\ref{fig:illustrate-method}(a) illustrates
the steps to obtain such a solution: First, two rough, fractal
surfaces are generated for each simulation, representing the two
sliding bodies. A fractal description is chosen because self-affinity
is often observed in experiments of rubbing surfaces
\citep{Majumdar1990, Persson2005, Persson2014, Davidesko2014,
  Candela2016}, as well as in simulations \citep{Milanese2019,
  Milanese2020a, Hinkle2020}. The surfaces are square and periodic in
$x$ and $y$ direction and represent half spaces in $z$
direction. Periodic synthetic surfaces are generated in the Fourier
domain with a fixed power spectrum and a uniform distribution of
phases~\citep{Wu2000}. Since all efficient contact algorithms describe
a rough-on-flat geometry, the model uses Johnson's assumption that the
rough-on-flat solution is equivalent to the rough-on-rough solution,
provided that the former is solved using the gap of the latter as the
rough surface~\citep{Johnson1985}.

Now, for a given contact solution at normal load $F_N$, all contact
junctions are identified using a flood-fill algorithm. Assuming that
the contacts are approximately circular, the critical length scale is
translated into a critical area $A^* = \pi {d^*}^2 / 4$ and junctions
are divided into plastically deforming and wear particle forming based
on their size (see Fig.~\ref{fig:illustrate-method}(b)). The idea is
now that Archard's wear coefficient indicates how many junctions can
form particles. The wear rate for each junction is obtained using
Archard's estimation that a junction fully detaches after a sliding
distance roughly equal to its diameter \citep{Archard1953}, therefore
obtaining a wear rate equal to volume divided by diameter, i.e.,
proportional to the junction's contact area. Note that this means that
the sliding distance needed to form a wear particle is different for
each junction. Archard's wear coefficient is thus calculated as
\begin{equation}
  k = \dfrac{\sum\limits_i
             \begin{cases}
               0   & A_i < A^* \\
               A_i & A_i \geq A^*
             \end{cases}}
            {\sum\limits_i A_i},
\end{equation}
where $A_i$ is the area of contact junction $i$ and the sum is over
all junctions. The wear rate is then simply $W = k \omega \sum_i A_i$,
where $\omega$ is an average shape factor on the order of unity that
accounts for the local contact and wear particle geometries. We call
this model ``static'' because the relative sliding motion is only
implied in the derivation of the per-junction wear rate.

\begin{figure*}
    \centering
    \includegraphics[center]{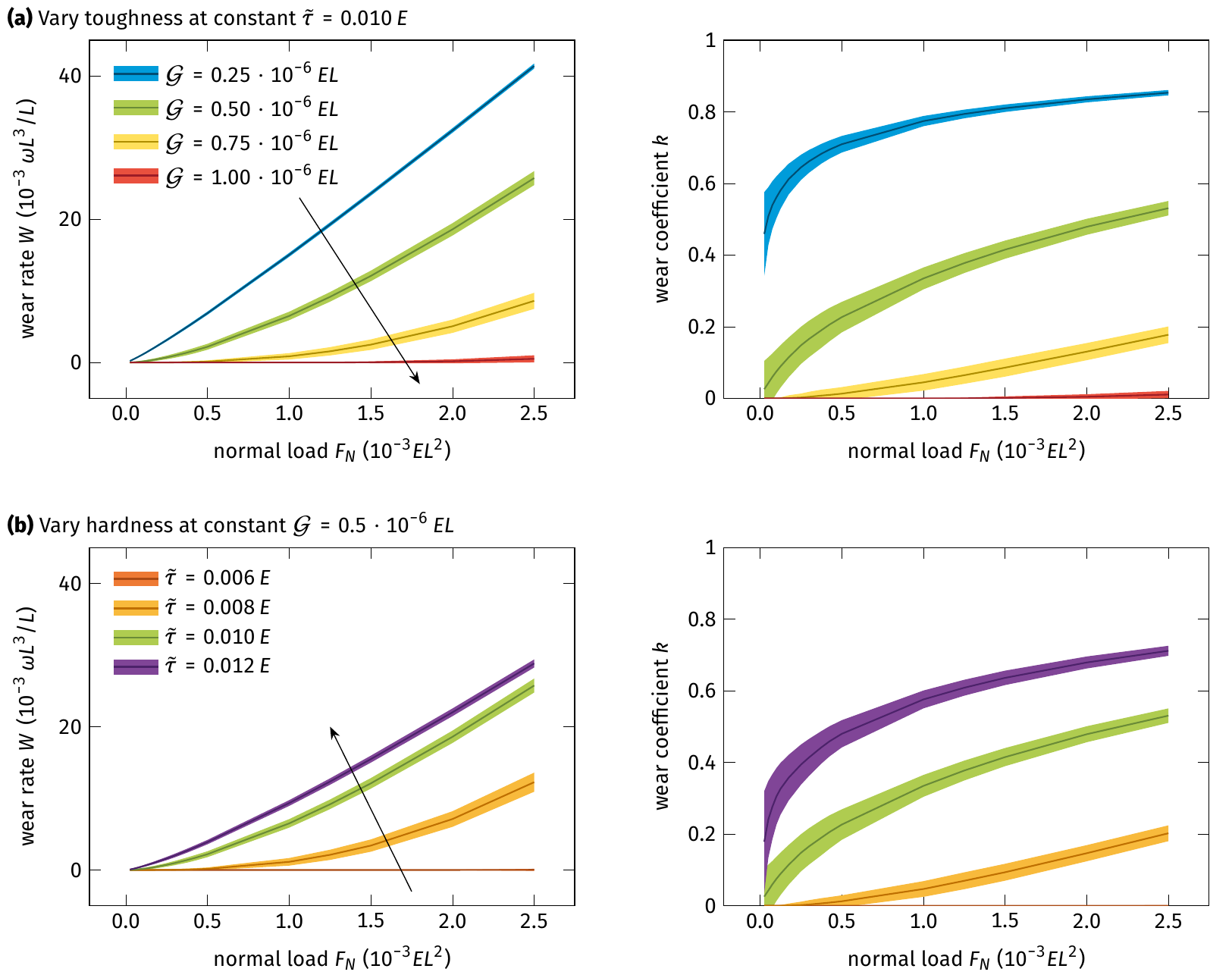}
    \caption{Wear rates and wear coefficients using the same model as
      \citet{Frerot2018} with the addition of a saturated-pressure
      contact solution instead of an elastic one. The wear rates are
      normalized by the shape factor $\omega$. The influence of
      varying (a) toughness and (b) hardness is shown. The colored
      area represents the standard deviation of the data between
      different simulation runs.}
    \label{fig:repeat-Frerot2018}
\end{figure*}

It is clear that a purely elastic contact solution is not realistic
\citep{Bowden1939,Weber2018} and our first step to improve upon the
above model is to introduce an elastoplastic contact solution. While
at least full von Mises plasticity is desirable for realistic contact
calculations, the required computing time is prohibitive for a large
number of simulations \citep{Pei2005, Frerot2019a}. In order to keep
the computational demands reasonable, we thus opted for a
saturated-pressure solution as described by \citet{Almqvist2007} and
as implemented in \textsc{tamaas} \citep{Tamaas, Frerot2020}. The
model relies on the common assumption that in elastic-plastic contact
the surface pressure should not exceed the hardness of the
material. To this end, saturated-plasticity approaches used in rough
contact~\citep[e.g.,][]{Akchurin2016,Weber2018} solve an elastic
contact problem with the additional constraint that the contact
pressure $p$ should be smaller or equal to $p_m$, the flow
pressure. Often, this flow pressure is taken to be equal to the
indentation hardness $\tilde{p}$, which can in turn be estimated from
the material's uniaxial yield strength $\tilde{\sigma}$ as
$\tilde{p} \approx 3\tilde{\sigma}$ under the assumption of spherical
indentation \citep{Tabor1951}. Note that this is an approximation of
the full elastoplastic contact solution and the readers are referred
to \citet{Frerot2020a} for a comparison between von Mises plasticity
and saturation pressure models.

We start our model development by investigating the influence of
plasticity on the wear prediction for the static wear model.  For this
set of simulations, we describe all quantities in reduced units in
terms of the Young's modulus $E$ and the side length $L$ of the
surface.  We used five realizations of each pair of rough surfaces,
each surface being discretized into $4096\times4096$ pixels and
described by a Hurst exponent $H = 0.7$ \citep[a typical value observed
experimentally, see][]{Persson2005}, a lower wavelength cutoff of
$\lambda_l = L/256$, an upper wavelength cutoff of $\lambda_u = L/16$,
and a root mean square (RMS) of heights of $h_\text{RMS} =
0.005\,L$. For each of those pairs, 1280 different offsets parallel to
the surface planes between the top and bottom surface were used, since
they were needed anyway for the sliding model described in
section~\ref{sec:sliding-model} later on.

We used an isotropic material with Poisson's ratio $\nu = 0.3$. For
the shear strength, values of
$\tilde{\tau} \in\{0.006,0.008, 0.010, 0.012\}\,E$ were chosen, from
which the flow pressure was derived as
$p_m = \tilde{p} = 3 \sqrt{3} \tilde{\tau}$. Values for the fracture
energy were $\mathcal{G} \in \{0.25, 0.5, 0.75, 1.0\}\cdot 10^{-6}\,EL$.
The geometry factor was set to $f = 3$ under the assumption of
spherical asperities, leading to a range of $d^*/L \approx 0.004$ to
$0.064$, i.e., in between $\lambda_l$ and $\lambda_u$. Sixteen
different normal loads were used, so that in total 409,600 contact
solutions were computed (the fracture energy only appears in $d^*$ and
has no bearing on the contact solution).

Despite the addition of more realistic contact physics, the results
resemble the work of \citet{Frerot2018}, see
Fig.~\ref{fig:repeat-Frerot2018}. The wear rate is superlinear and the
wear coefficient increases with load, tending towards a limit of
one. Additionally, it should be noted that while the wear rate
increases with decreasing fracture energy (first row of
Fig.~\ref{fig:repeat-Frerot2018}), which seems reasonable, it also
increases with increasing hardness (second row of
Fig.~\ref{fig:repeat-Frerot2018}), which contradicts Archard's wear
law. This is the same in the original work and is a simple consequence
of using $d^*$ as a wear particle formation criterion: The bigger
$d^*$, the fewer junctions on a given contact map form wear
particles. Since $d^* \propto \tilde{\tau}^{-2}$, this effect cannot
be offset by the increasing real contact due to plasticity, which is
less than quadratic with $\tilde{\tau}$~\citep{Bowden1939}.

The intermediate conclusion is either that the critical length scale
model contradicts experimental evidence or that a wear model based on
a single, static contact solution is missing a critical ingredient. We
will argue for the latter by implementing a simple sliding process.

\section{Sliding Model}
\label{sec:sliding-model}

In order to explicitly include the evolution of wear particle
formation during the sliding process in our simulations, we extended
the model presented in the previous section by continuously displacing
the top and bottom surfaces against each other. We do not calculate
wear rates directly from the real contact areas, but instead estimate
wear volumes under the assumption that the junctions are circular and
the resulting wear particles spherical (take $A_i$ as the area of the
contact spot $i$ and use $d_i \approx 2\sqrt{A_i/\pi}$, resulting in a
wear particle volume $v_i \approx \pi d_i^3/6$). The algorithm---also
shown in Fig.~\ref{fig:illustrate-method}(c)---works as follows:
\begin{enumerate}
\item Solve the contact, find all contact junctions, and record the
  instantaneous wear volume $V_0$ as
  \begin{equation*}
    V_0 = \sum_i
    \begin{cases}
      0   & d_i < d^* \\
      v_i & d_i \geq d^*
    \end{cases},
  \end{equation*}
  where the sum goes over all contact junctions $i$, which were
  identified using a flood-fill algorithm.
\item \label{enum:algo:displace} Displace the top surface by a sliding
  increment $\delta s$ and recompute the contact solution. Due to the
  periodic boundary conditions, the part of the top surface that
  leaves the simulation box on one side is in fact reinserted on the
  opposite side.
\item Find all contact junctions. Iterate over them and mark any
  junction that shares at least one pixel with a junction that has
  formed a wear particle in the previous step as ``disabled''. The
  same for any junction that shares at least one pixel with a junction
  that was in the disabled state in the previous step. All other
  junctions are marked ``active''.

  Thus, a junction (which represents an asperity--asperity contact)
  goes through several phases: When it first appears, it is
  approximately a single point and only grows during further
  sliding. Should it reach a size of $d > d^*$ before disappearing, a
  wear particle is formed and the junction is marked as disabled
  afterwards. It will retain its disabled state until the junction
  shrinks and disappears, at which point its state is no longer
  stored.
  This mimics the ongoing detachment of a wear particle, whose volume
  should only be counted once, but which still contributes to carrying
  normal load. Because the long-distance rolling of a wear particle
  cannot be modeled in the present framework, we assume the particle
  is lost quickly.
\item Increment the wear volume by
  \begin{equation*}
    \delta V = \sum_i
    \begin{cases}
      0   & d_i < d^* \text{ or disabled} \\
      v_i & d_i \geq d^* \text{ and active.}
    \end{cases}
  \end{equation*}
\item Unless the desired total sliding distance was reached, go to step
  \ref{enum:algo:displace}.
\end{enumerate}
The output of this algorithm is a cumulative wear volume as a function
of sliding distance. We chose a total sliding distance of
$L/2$. \hyperref[sec:suppl-data]{Video~1} shows an example simulation
of a smaller surface. The surface size was reduced to make the
features of the model more easily visible.

\begin{figure}
  \centering
  \fcapside[9cm]{%
    \caption{Running in and fitting the wear rate. The wear rate is
      the slope obtained by linear regression. The initial points
      indicated in the inset correspond to the instantaneous wear
      volume $V_0$, which is much higher than any subsequent increment
      $\delta V$. Material parameters used for these simulations were
      $\tilde{\tau} = 0.012\,E$ and
      $\mathcal{G} = 0.5\cdot 10^{-6}\,EL$.}
    \label{fig:running-in}%
  }{%
    \includegraphics[]{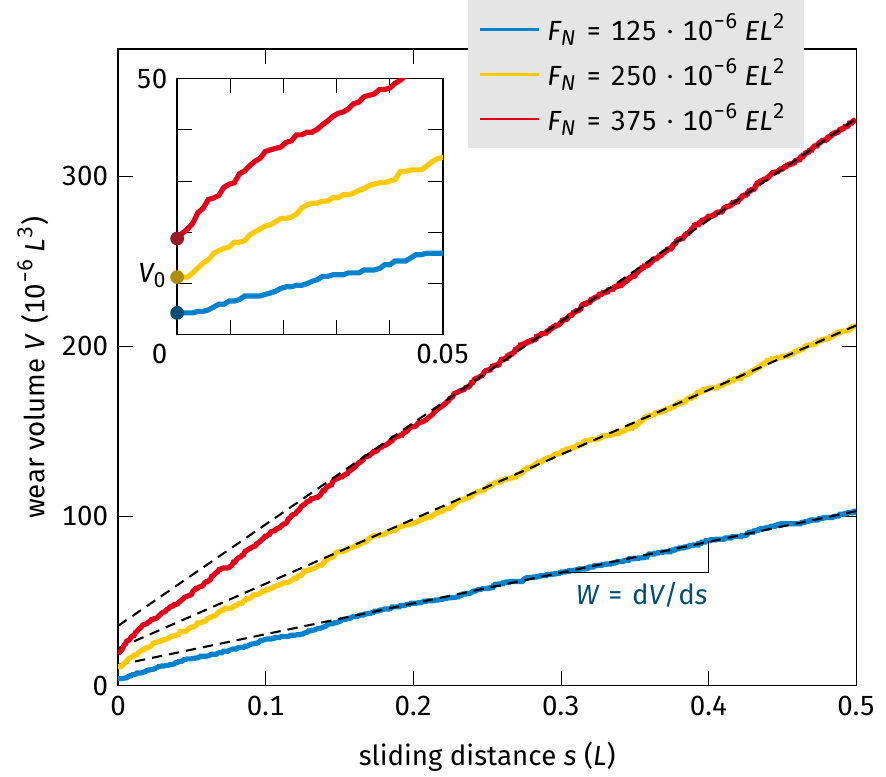}%
  }
\end{figure}

Figure~\ref{fig:running-in} shows that the wear volume increases
linearly with the sliding distance after a running-in period, in which
the increase can be steeper. As in virtually all experiments
\citep{Reye1860, Rabinowicz1951, Burwell1952, Archard1953,
  Archard1956}, our wear rate prediction (obtained through linear
regression as $W = \mathrm{d}V/\mathrm{d}s$) reaches a constant value,
yielding a first confirmation that the sliding implementation is
reasonable. Note also that the instantaneous wear volume $V_0$ is much
higher than the increments $\delta V$. We expect this effect to be
very hard to observe experimentally, since it takes place at the exact
onset of sliding. Nevertheless, it has an analogy in the
well-established observation that friction at rest is increased
compared to friction during sliding. As can be seen in
Fig.~\ref{fig:running-in}, the system retains a ``memory'' of that
initial contact and the wear rate stabilizes only after a while. Such
running-in periods with higher wear volume are well known from
experiment \citep{Queener1965}, although they can have many causes not
included in our model, such as changes of the surface roughness
parameters. A measure of this memory in the present work is the number
of remaining contacts that were present at $s = 0$, which we found to
decay exponentially with sliding distance \citep[as reported before
by][]{Dieterich1994, Pei2006PhD}. We could, however, not find a way to
predict the running-in distance: The decay constant is not uniform for
different surfaces or normal loads and there is no relation between
the size of the largest contacts and running in, as one might have
suspected.

For the following study, we thus defined a sliding distance of $L/4$
as a safety margin after which the system is definitely run in, which
we confirmed by inspecting the simulation data. This running-in
distance has no physical meaning. We chose a sliding increment
$\delta s$ corresponding to 4~pixels in the running-in period and
1~pixel afterwards.\footnote{We tested several, uniform values of
  $\delta s$ for the whole sliding distance, but found that changes in
  slope, i.e., wear rate, due to large steps occur quickly. Increasing
  the step size only in the running-in phase did not however influence
  the slope after running in up to a running-in step size
  corresponding to 16~pixels. We thus settled on the present values as
  sufficient safety margins, while still accelarating the simulations
  significantly.}

Additionally, we verified that the ratio $\lambda_l/\Delta x$, where
$\Delta x$ is the side length of a pixel, gives an acceptable
discretization error. The details can be found in
\ref{sec:discretization}.

\subsection{Role of the critical length scale}

The critical length scale $d^*$ provides a minimum size for a junction
to form a wear particle and therefore a minimum wear particle volume
$v^* \approx \pi {d^*}^3/6$. The result in the static model of
section~\ref{sec:repeat-Frerot} is a variety of wear particle sizes
$v_i \geq v^*$ (see, e.g., Fig.~\ref{fig:illustrate-method}(b)) and a
strong dependence of the wear coefficient on the maximum wear particle
size \citep{Frerot2018}. In the initial step of our simple sliding
model, this is the same and we obtain a relatively high instantaneous
wear volume $V_0$ with a large number of wear particles of different
sizes. The ongoing sliding process, however, changes the distribution
of particle sizes. All junctions with an initial size $d > d^*$ are
already worn off. During sliding, more asperities come into contact
and their contact junctions grow with~$s$. As soon as they reach a
size of $d^*$ (if they do not shrink again prior to reaching it) they
immediately form a wear particle of volume $v^*$ before they can grow
bigger.  That means that the wear volume increment $\delta V$ is made
up of wear particles of volume $v^*$. This reveals the role of the
critical length scale as not only the minimum size of wear particles,
but probably also the common size. Thus infinite particle sizes, as
featured in the work of \citet{Popov2018a}, do not occur in reality
because contact junctions simply wear off before they can grow that
big. It is known from experiments, e.g., by \citet{Rabinowicz1953,
  Rabinowicz1995} or more recently by \citet{Kirk2020}, that there is
a distribution of particle sizes. This can have two reasons. On one
hand, $d^*$ is not uniquely defined on a real surface due to
fluctuations of the adhesion and local geometry, which are both
expected to affect $d^*$ \citep{Brink2019}. On the other hand, once
particles have detached, they continue to evolve and grow, losing the
connection to their initial size \citep{Milanese2019,
  Aghababaei2019a}. Experimentally, even sintering of particles has
been observed \citep[see, e.g.,][]{Kirk2020}.

Given that the contribution of the initial particle creation to the
wear rate is large compared to the particle growth rate
\citep{Milanese2019}, we neglect any wear due to particle growth. The
wear rate can thus be defined by both $d^*$ and the rate at which new
contact junctions appear and reach $d = d^*$ during sliding:
\begin{equation}
    \label{eq:wear-rate-new}
    W = \omega {d^*}^3 \frac{\mathrm{d}}{\mathrm{d}s}N_+^*(s, d^*, F_N).
\end{equation}
Here, $\omega$ ($=\pi/6$ in the present work) is again a shape factor
accounting for the wear particle shape and $N^*_+$ is the total number
of wear particles formed during the sliding distance $s$, which is
also the cumulative number of junctions that have newly grown to a
size of at least $d^*$. This number $N^*_+$ should be a linear
function of $s$ to obtain a constant wear rate. Note that the total
number of junctions greater than $d^*$ at a given sliding distance,
denoted $N^*$, is roughly constant after running in. For more details,
see \ref{sec:number-of-wear-particles}.

\subsection{Apparent contact area}
\label{sec:model:apparent-contact-area}

Typically, it is assumed that the wear rate, just like the real
contact area, is independent of the apparent contact area
\citep{Archard1953}. We tested if this is the case in our model (see
\ref{sec:apparent-contact-area} for the details) and found for two
surfaces with different side lengths that the wear rates are
comparable up to a given load, after which the smaller surface
exhibits a breakdown of the wear rate with increasing load. This is
not related to a breakdown of the total number of contact
junctions---which strongly depends on the surface size---but to a
breakdown of the scaling of $N^*$ with $F_N$, i.e., the number of
contact junctions greater than $d^*$. This seems to have direct
implications on $\mathrm{d}N_+^*/\mathrm{d}s$ and thus the wear rate.
Here, we treat this as an effect of the limited size of our surfaces
and regard the data after the ``breakdown'' as unreliable. If such a
limit to Archard-like scaling with $F_N$ is a real effect remains to
be investigated.

\subsection{Lower wavelength cutoff and the RMS of heights}

In elastic contact solutions, the real contact area scales linearly
with the inverse of the RMS of slopes \citep{Bush1975,
  Hyun2004}. Because the latter is strongly dependent on $\lambda_l$,
the true contact area is sensitive to local fluctuations at short
wavelengths and \emph{in extenso} to the resolution down to which the
system is measured or modeled \citep{Ciavarella2000, Persson2001a,
  Jacobs2017aComment}.  This is a problem for elastic contact
simulations, since it becomes necessary to include the shortest
wavelength of roughness that exists in the system, which might be much
smaller than $d^*$ and which is likely unknown due to experimental
measurement limitations \citep{Jacobs2017}. However, in elastoplastic
contact solutions, plasticity can ``smooth out'' small wavelengths
\citep{Pei2005}. We found that this also is the case in our model, see
\ref{sec:lower-wavelength-cutoff}. While it is hard to generalize to
other elastoplastic models, it at least means that the present model
is robust against (in practice) immeasurable small disturbances of the
rough surfaces.

\begin{figure}[t!]
    \centering
    \includegraphics[center]{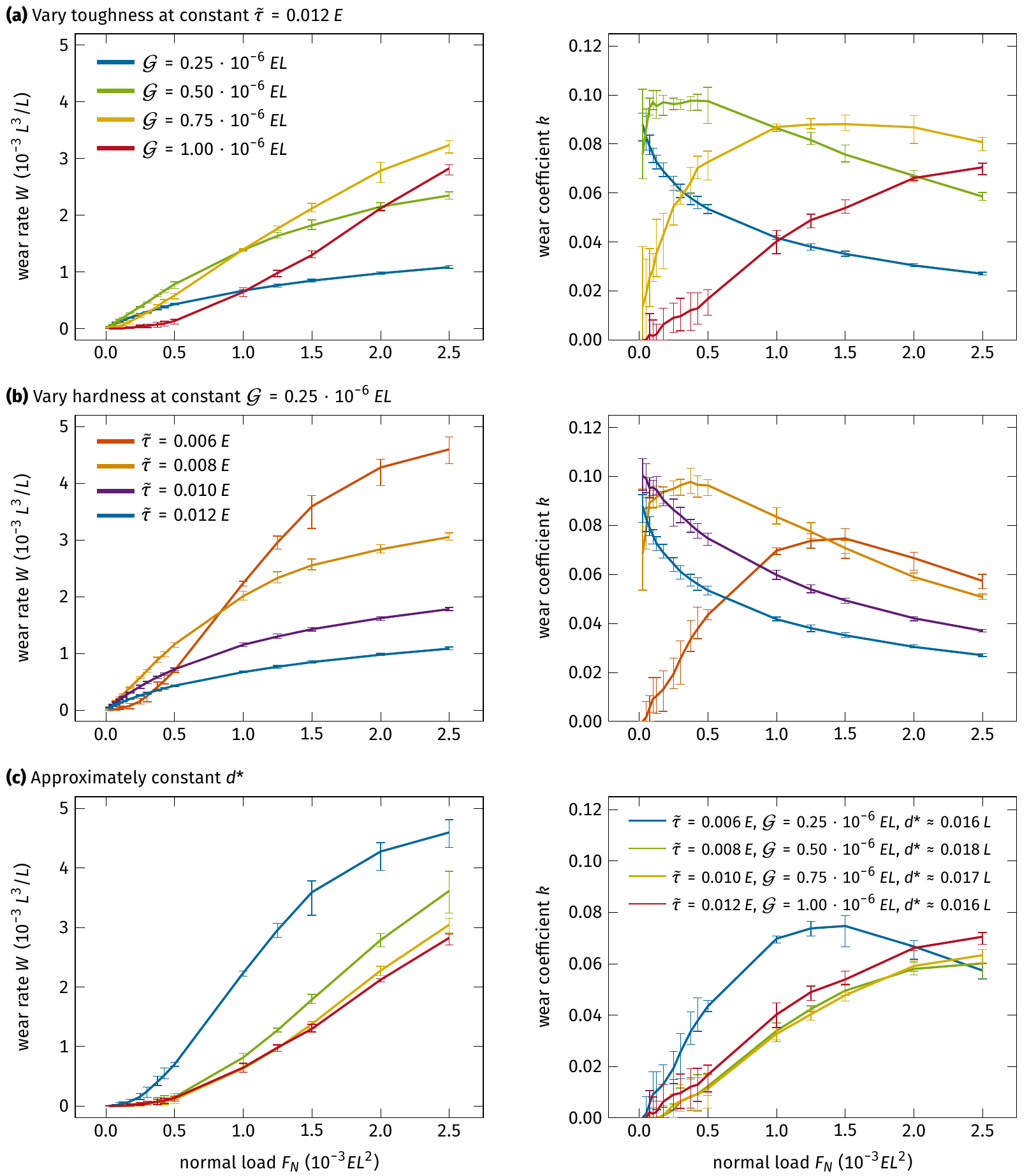}
    \caption{Wear rates and wear coefficients predicted by our
      model. Each data point is the result of five independent, full
      sliding simulations over a sliding distance $s=0.5\,L$ as shown
      exemplarily in Fig.~\ref{fig:running-in}. The data is the
      average of the five simulations, the error bars indicate the
      minimum and maximum values. Wear coefficients were calculated
      using Eq.~\ref{eq:Archard-law}. Critical length scales $d^*$ in
      order of appearance in the legend are approximately (a)
      $0.004\,L$, $0.008\,L$, $0.012\,L$, $0.016\,L$; (b) $0.016\,L$,
      $0.009\,L$, $0.006\,L$, and $0.004\,L$. Wear rates at higher
      loads are usually higher for materials with larger $d^*$.}
    \label{fig:wearrate-reduced}
\end{figure}

Another surprising property is that the wear rates do not change
between two values of the RMS of heights $h_\text{RMS}$ tested by us
(see \ref{sec:h-rms}). For an elastic contact solution this would be
unexpected, since the RMS of heights also influences the RMS of slopes
and thus the contact solution. In our case, it seems that the nature
of the sliding model combined with the saturated-pressure contact
solution cancels any influence of $h_\text{RMS}$, at least in a
certain range of $h_\text{RMS}$ and $F_N$. This can be rationalized by
first taking into account that sharper peaks also increase the local
pressure and thereby provide the necessary energy to flatten them, as
well as the fact that the wear rates in our model depend more strongly
on the number of contact junctions, which is likely not affected as
strongly by $h_\text{RMS}$. Further study is necessary to draw
definite conclusions.

\begingroup
\setlength{\textfloatsep}{18pt plus 2 pt minus 4pt} %
\begin{figure}
    \centering
    \includegraphics[center]{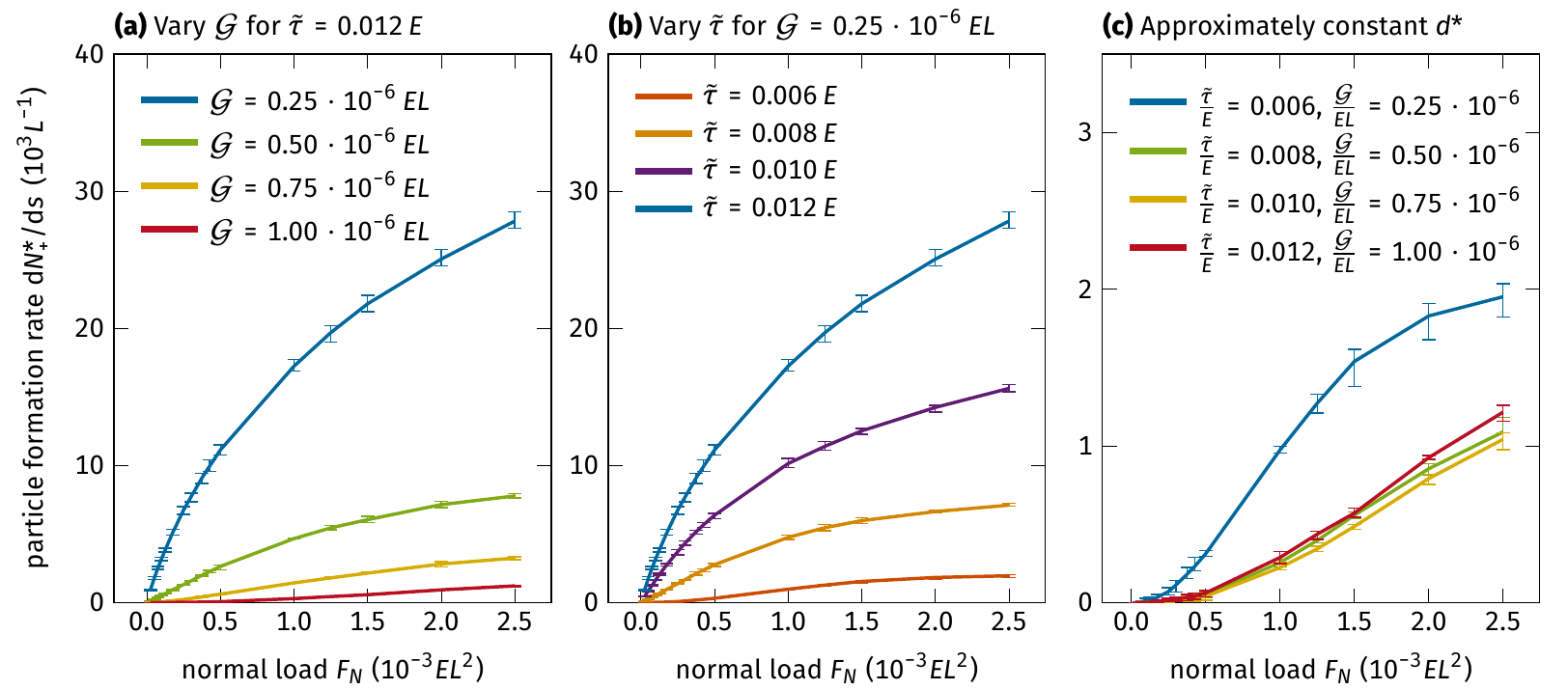}
    \caption{Wear particle formation rates
      $\mathrm{d}N^*_+/\mathrm{d}s$ for the simulations depicted in
      Fig.~\ref{fig:wearrate-reduced}. The meaning of the error bars
      is equivalent. Note the inversions in (a) and (b): The hardest
      material and the least tough material wear least, but form the
      most individual wear particles. They are just much smaller.}
    \label{fig:dNplusds}
\end{figure}

\section{Results}

\subsection{Influence of material properties}
\label{sec:res:red-units}

We used the surfaces described in section~\ref{sec:repeat-Frerot} as
input to our sliding model with identical variations of material
parameters. The results are shown in Fig.~\ref{fig:wearrate-reduced}
and \hyperref[sec:suppl-data]{Supplementary} Figs.~S.1 and S.2. We
observe that wear rates and wear coefficients are an order of
magnitude smaller than in the static model
(Fig.~\ref{fig:repeat-Frerot2018}). They reside in the high end of
reported experimental values for dry sliding experiments of ceramics
\citep{ZumGahr1989, Hokkirigawa1991, Wang1996} or of steel on steel
\citep{Archard1956}, which range from $k = 10^{-6} $ to $10^{-1}$.
Unlike the results shown in section~\ref{sec:repeat-Frerot}, the wear
coefficient here does not asymptotically tend to unity, but remains
smaller than around $0.1$ in all cases. More insights can be gained by
exploring the influence of the material parameters.

First, we kept the material hardness constant and varied the fracture
energy $\mathcal{G}$ (Fig.~\ref{fig:wearrate-reduced}(a)). This means
that the distribution of contact patches remains the same for these
simulations, since the contact solution does not depend on the
fracture energy. However, the wear particle formation criterion is
affected, because $d^* \propto \mathcal{G}$. At low loads, the systems
with smaller fracture energy wear more. This is due to the same
mechanism we have discussed in section~\ref{sec:repeat-Frerot} with
regards to the static model: When $d^*$ is large, most junctions that
form between the two surfaces are too small to be able to detach as
wear particles. When the load is increased, more and more large
junctions occur. If a sufficient number of junctions grow to a size of
$d^*$, the larger wear particle volume $v^* \propto {d^*}^3$ of the
tougher materials ultimately makes them wear more than the less tough
ones. Nevertheless, the increase of the wear coefficient with $F_N$ is
slower for tougher materials and they reach their maximum value of $k$
at higher $F_N$. This slow increase of the wear coefficient is
exemplified by the simulation with $\mathcal{G} = 10^{-6}\,EL$, whose
wear coefficient is still increasing as a function of the normal load
at $F_N = 2.5 \cdot 10^{-3}\,EL^2$ and has not yet surpassed the
simulation with $\mathcal{G} = 0.75 \cdot 10^{-6}\,EL$ in wear rate,
even though it has a larger $d^*$.

Second, by keeping the fracture energy constant and varying the
material strength $\tilde{\tau}$, we can observe the effects of
simultaneously changing the contact solution---which is a function of
the indentation hardness $\tilde{p} \propto \tilde{\tau}$---and the
critical length scale $d^* \propto \tilde{\tau}^{-2}$
(Fig.~\ref{fig:wearrate-reduced}(b)). Again, larger $d^*$ leads to
higher wear rates at high loads. As before, we would expect that the
systems with large $d^*$ wear less at low loads. However, this is only
the case for the simulation with the lowest hardness, making the
effect less pronounced than in the previous set of simulations. The
reason is the interplay between $d^*$ and the contact solution:
Reducing the material strength decreases the indentation hardness and
leads to larger contact junctions even at low loads, counteracting the
increase of $d^*$.

Finally, by keeping $d^*$ approximately constant and varying
$\tilde{\tau}$ and $\mathcal{G}$ together
(Fig.~\ref{fig:wearrate-reduced}(c)), we see that the wear
coefficients approach the same maximum value of
$k \approx 0.06$--$0.07$. Again, stronger and tougher materials reach
this maximum at somewhat higher normal loads.

It is also instructive to look at the number of wear particles, or
rather the rate at which they are emitted over a given sliding
distance: $\mathrm{d}N^*_+/\mathrm{d}s$. This is shown in
Fig.~\ref{fig:dNplusds} for the same set of simulations. It should be
noted that while the hardest material and the material with the lowest
fracture toughness wear the least, they emit the most wear
particles. Since these particles are quite small in these cases,
though, the wear rates are also small. This means that more severe
wear is due to the detachment of large wear particles, which is also
sometimes observed experimentally \citep{Zhang1997}. Of course, when
$d^*$ is roughly constant, the wear particle formation rate
(Fig.~\ref{fig:dNplusds}(c)) and the wear rate
(Fig.~\ref{fig:wearrate-reduced}(c)) exhibit similar trends. Thus, if
one wants to minimize the emission of fine particles, our current
model suggests to use a soft but tough material.

Experiments often exhibit wear coefficients at least one or two orders
of magnitude below the present results \citep{Archard1956,
  ZumGahr1989, Hokkirigawa1991, Wang1996}, which is likely due to
reduced adhesion even in dry experimental conditions. We assumed in
this work that the junctions have bulk strength, although surface
contamination is always present and will reduce the adhesion between
the two bodies in contact. Indeed, it is well known that friction in
vacuum is increased and that this is further amplified upon wearing
off of surface contaminants, although the effect is smaller for brittle
materials (such as ceramics) than for metals \citep{Kato1990,
  ZumGahr2001}. The reduced adhesion in normal atmosphere would lead to
an increased $d^*$ \citep{Brink2019, Aghababaei2019a}, while not
affecting the bulk hardness and contact solution. This is equivalent
to increasing $\mathcal{G}$ while keeping all other material
parameters the same (see Eq.~\ref{eq:d-star}).
Figure~\ref{fig:wearrate-reduced}(a) shows that this leads to a
reduction of wear compared to simulations with smaller $d^*$ at low
loads, since there are less junctions with size $d > d^*$ at low
$F_N$. By significantly reducing adhesion (possibly by lubrication),
even the onset of wear can be delayed to very high loads in the
present model, see for example
\hyperref[sec:suppl-data]{Supplementary} Fig.~S.1(a). As noted in
\citet{Brink2019} and \citet{Milanese2020a}, though, the critical
junction size becomes geometry dependent in the reduced adhesion case
and some wear particles \emph{can} be formed, thereby resulting in a
reduced, but non-zero wear coefficient. Unfortunately, the actual
microscopic junction shear strength for reduced adhesion is not easily
accessible experimentally and the above hypothesis remains to be
tested against realistic data.

\endgroup %

\subsection{Comparison to data from Kim et al.\ (1986)}

A direct comparison to experimental data is not easy, since we could
not find any work which simultaneously reports the surface roughness
parameters, material properties, and wear rates for dry sliding
conditions. Nevertheless, we could use the data for Si$_3$N$_4$, SiC,
and Al$_2$O$_3$ from \citet{Kim1986}, who at least report the RMS of
heights for their surfaces in addition to material properties and wear
rates (see Table~\ref{tab:mater-prop-Kim1986}), even if only for
rolling friction experiments.

In terms of the surface geometries, we used a side length of
$L_K = \SI{200}{\micro\meter}$ (roughly the apparent contact area of
the experiment, although their contact area is an ellipse due to
Hertzian contact); $H=0.7$,
$\lambda_l = L_K/512 = \SI{0.39}{\micro\meter}$, and
$\lambda_u = L_K/8 = \SI{25}{\micro\meter}$. As such, $d^*$ lies in
between $\lambda_l$ and $\lambda_u$, keeping the computational demands
(discretization) reasonable. The RMS of heights was chosen to be the
steady-state value of the wear experiment (see
Table~\ref{tab:mater-prop-Kim1986}). Note that the values for
$\tilde{\tau}/E$ and $\mathcal{G}/EL_K$ lie in the same range as the
values we used for our generic simulations with reduced units,
verifying that the choice of parameters was reasonable, at least for
ceramics.

\begin{table}[b]
    \centering
    \caption{Material and surface properties of several ceramics. Data
      from \citet{Kim1986}. All synthesized model surfaces had a side length of
      $L_K = \SI{200}{\micro\meter}$, Hurst exponent of $H = 0.7$,
      $\lambda_l = \SI{0.39}{\micro\meter}$, and
      $\lambda_u = \SI{25}{\micro\meter}$. The reduced units lie in
      the same range as before.}
    \label{tab:mater-prop-Kim1986}
    \begin{tabular*}{\textwidth}{p{1.6cm}@{\extracolsep{\fill}}cccccccc}
        \toprule
        Material & $E$ (GPa) & $\nu$ & $\tilde{\tau}$ (GPa) & $\tilde{\tau}/E$ & $\mathcal{G}$ (J/m$^2$) & $\mathcal{G}/EL_K$ & $d^*$ (\si{\micro\meter}) & $h_\text{RMS}$ (\si{\micro\meter}) \\
        \midrule
        Si$_3$N$_4$ & 294 & 0.27 & 2.89 & 0.0098 & 91.97 & $1.56\cdot10^{-6}$ & 7.64 & 4.0 \\
        SiC         & 392 & 0.16 & 4.80 & 0.0122 & 45.00 & $0.57\cdot10^{-6}$ & 1.98 & 5.0--10.0 \\
        Al$_2$O$_3$ & 343 & 0.25 & 3.37 & 0.0098 & 46.65 & $0.68\cdot10^{-6}$ & 3.93 & 20.0 \\
        \bottomrule
    \end{tabular*}
\end{table}

\begin{figure}
    \centering
    \includegraphics[center]{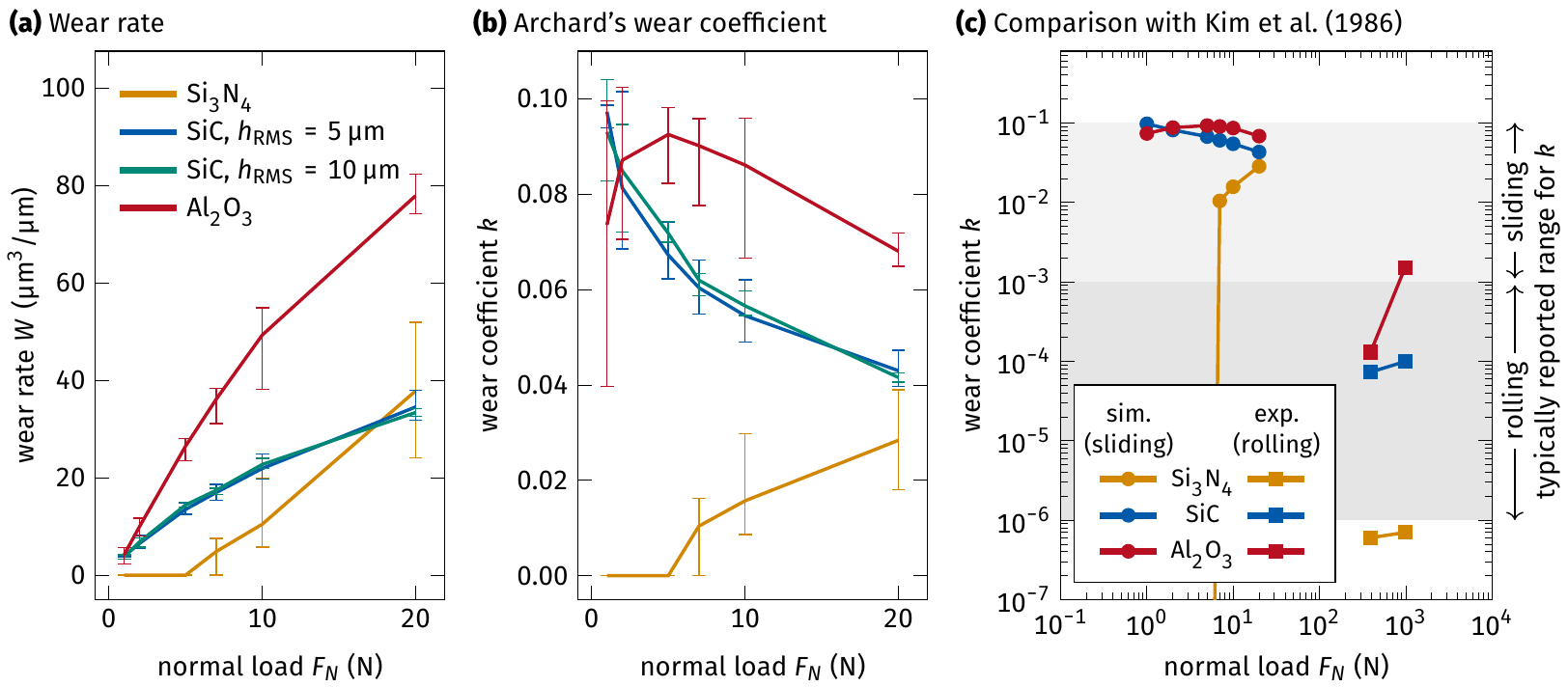}
    \caption{Results of the simulations using material parameters
      similar to the experiments of \citet{Kim1986}. (a) Wear rates
      and (b) wear coefficients are shown. (c) A comparison between
      the sliding simulations and the rolling experiments shows the
      same hierarchy of wear severity between both, but
      large differences in the absolute values. The shaded areas
      indicate typical ranges for sliding and rolling wear experiments
      as reported by \citet{Hokkirigawa1991}. The large absolute
      differences between simulation and experiment are thus expected
      due to the different setup.}
    \label{fig:wearrate-Kim1986}
\end{figure}

The results are plotted in Figs.~\ref{fig:wearrate-Kim1986}(a)--(b)
and the wear coefficients are in the same range as the previous
simulations (Fig.~\ref{fig:wearrate-reduced}), as expected. The
authors of the experimental work proposed a wear law based on
``contact severity'', which does not seem to yield insights beyond
Archard's wear law. It is discussed for completeness's sake in
\ref{sec:Sc}. Apart from this, Fig.~\ref{fig:wearrate-Kim1986}(c)
shows that our wear coefficients are at least two to three orders of
magnitude higher than the experimental ones
($k_\text{exp} = 6\cdot10^{-7}$ to $1.5\cdot10^{-3}$), although this
is expected since the experiment was a rolling friction setup, which
often has such a large difference in $k$ compared to sliding friction
setups \citep{Hokkirigawa1991}. The hierarchy of wear resistance
matches the experimental observation: Al$_2$O$_3$ wears the most and
Si$_3$N$_4$ the least for the presented load range. While the
crossover between Al$_2$O$_3$ and SiC at low loads observed in the
simulation is also somewhat suggested by the experimental trends, the
available measurement data is insufficient to affirm a
correspondence. In contrast, the wear rate of Si$_3$N$_4$ predicted by
our model increases towards the higher loads in accordance with the
results of section~\ref{sec:res:red-units}, since it has the highest
$d^*$. It is unclear if that behavior of our model is realistic or
not, since we cannot translate the load ranges in sliding to the
equivalent load ranges in rolling friction. It is quite possible that
the typical experimental load ranges never reach the point where the
wear rate of Si$_3$N$_4$ picks up and surpasses the other materials.
Some hints can be found in the sliding wear experiments on different
ceramics of \citet{Adachi1997}. They report wear rates up to the order
of \SI{10}{\micro\meter^3/\micro\meter} in dry conditions for normal
loads in the range of \SI{2}{N} to \SI{100}{N}, which corresponds
roughly to our results. Instead of normal loads, data is presented as
a function of ``contact severity'' (a compound value of different,
independent parameters), which makes further comparison impossible.

Ultimately, we hope to inspire more detailed experiments that
adequately report on material parameters, surface roughness, adhesion
(preferably at the individual junction level), and wear rates over
large load ranges in order to verify and improve our model. Recent
advances in the mass resolution and wear mapping of sliding
experiments, as for instance proposed by \citet{Garabedian2019}, can
be helpful in this context because they allow observations at the
onset of wear and can resolve asperity-level mechanisms.

\section{Outlook on wear modeling}

Due to the computational demands, the present model is relatively
simple. It is clear that a saturated-pressure contact solution is not
as satisfactory as a more realistic von Mises plasticity
\citep{Frerot2020a}, but existing approaches \citep{Pei2005,
  Frerot2019a} are computationally too demanding for the large number
of contact solutions required at present. Similarly, the spectrum of
the fractal roughness is not very broad, once again to keep
computational demands low. Finally, Johnson's assumption is likely
invalid for elastoplastic contact and true rough-on-rough contact
algorithms should be developed. Mechanistic wear modeling could
greatly benefit from any advances in these areas, if only to verify
the validity of the approximations used here.

There are also several physical processes that are not explicitly
included. First, a reduced adhesion between the surfaces, either due
to surface contamination or lubrication, is known to modify $d^*$
\citep{Brink2019}. However, a purely geometric treatment of $d^*$ as
used here would additionally need the contact angles
\citep{Brink2019}, which are not well defined when using Johnson's
assumption. Perhaps it would be more fruitful to work on including the
fracture process directly in the simulation instead, as attempted by
\citet{Frerot2020a}. Moreover, coupling of a fracture model to the
contact solution could inherently include asperity interactions
\citep{Aghababaei2018, Pham-Ba2020}, which are expected to greatly
affect the wear rates at high loads. Finally, we assumed that the
debris particles are somehow evacuated quickly and no extended third
body forms, which is of course an oversimplification in many cases.

Despite all of this, a major advantage of the current work is the
simplicity of Eq.~\ref{eq:wear-rate-new}. The main complication is to
provide a reliable estimate of $N^*_+$ or
$\mathrm{d}N^*_+ / \mathrm{d}s$. If a closed form solution could be
obtained, the computational demand would be significantly reduced and
the model could be widely and cheaply tested and applied.

\section{Conclusion}

We developed a wear model based on a combination of a simple
elastoplastic contact solution algorithm, an evolution of the contact
interface through a sliding process, and a critical length scale
criterion for wear particle formation. The model contains no fit
parameters and yields at least qualitatively reasonable results. We
find that the critical length scale $d^*$, i.e., the minimum size a
contact needs to have to detach and form a wear particle, is a
necessary ingredient in this model: It defines not only a minimum wear
particle size, but rather a typical wear particle size. Often, this
leads to materials with larger $d^*$ wearing more, which are materials
that are soft but tough (see Eq.~\ref{eq:d-star}), thereby reconciling
the concept of the critical length scale with Archard's wear law. For
smaller loads increased toughness can suppress wear particle formation
since no large enough contact junction forms, making tough materials
more wear resistant in some load ranges. The statistical distribution
of contact junctions that form during sliding is thus also a relevant
parameter. It remains to be seen if the behaviors at different load
ranges can be related to experiments.

Nevertheless, an Archard-like wear behavior is approximately recovered
and the qualitative agreement with an experimental work is
promising. However, the present comparison is not quantitatively
satisfying, since we can only compare to rolling wear, while we
intrinsically model sliding wear, for which we could not find
sufficient data. It is our hope that such modeling inspires more
detailed dry, adhesive wear experiments, which report simultaneously
the necessary material parameters, surface roughness description after
running in, and resulting wear rates. Preferably the work would be
performed over large load ranges in order to evaluate and improve
models such as the one presented in the present paper.

\section*{CRediT author contribution statement}

\textbf{Tobias Brink:} Conceptualization, Methodology, Investigation,
Writing -- Original Draft, Visualization. \textbf{Lucas Fr\'erot:}
Methodology, Software, Writing -- Review \&
Editing. \textbf{Jean-Fran\c{c}ois Molinari:} Conceptualization,
Writing -- Review \& Editing, Supervision, Funding acquisition.

\section*{Acknowledgments}

This work was supported by {\'E}cole polytechnique f\'ed\'erale de
Lausanne (EPFL) through the use of the facilities of its Scientific IT
and Application Support Center.

\appendix

\section{Discretization}
\label{sec:discretization}

\begin{figure}[h]
  \centering
  \fcapside[9cm]{%
    \caption{Verification of the discretization. The simulations were
      run 5 times each, the error bars indicate the minimum and
      maximum value of the wear rate.}
    \label{fig:discretization}
  }{%
    \includegraphics[center]{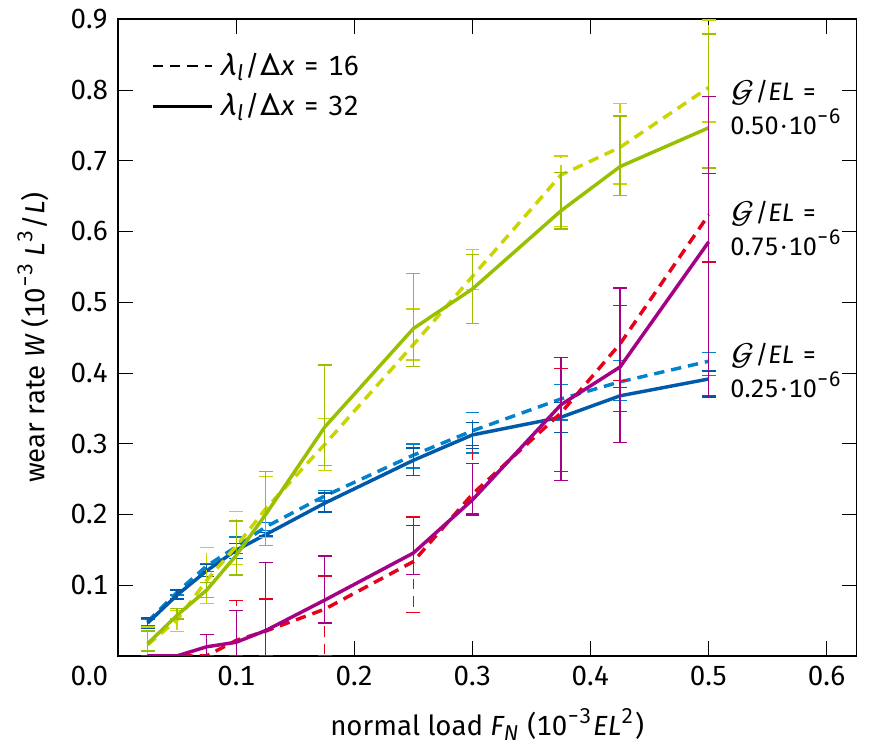}%
  }
\end{figure}

In order to verify that the discretization errors are sufficiently
small, we ran simulations with a ratio $\lambda_l/\Delta x = 16$ as
used in the results in this paper, as well as
$\lambda_l/\Delta x = 32$, i.e., a two times finer discretization. All
test simulations, presented in Fig.~\ref{fig:discretization}, used
roughness parameters of $H = 0.7$, $\lambda_l = L/256$,
$\lambda_u = L/16$, and $h_\text{RMS} = 0.005\,L$, as well as material
properties of $\nu = 0.3$, $\tilde{\tau} = 0.010\,E$ and varying
$\mathcal{G}$ as indicated in the figure.  We found no difference
between the two settings within the error of the simulation and thus
opted for a value of $\lambda_l/\Delta x = 16$ in the interest of
reducing computational demand due to discretization.

\section{Number of wear particles}
\label{sec:number-of-wear-particles}

\begin{figure}[b]
    \centering
    \includegraphics[center]{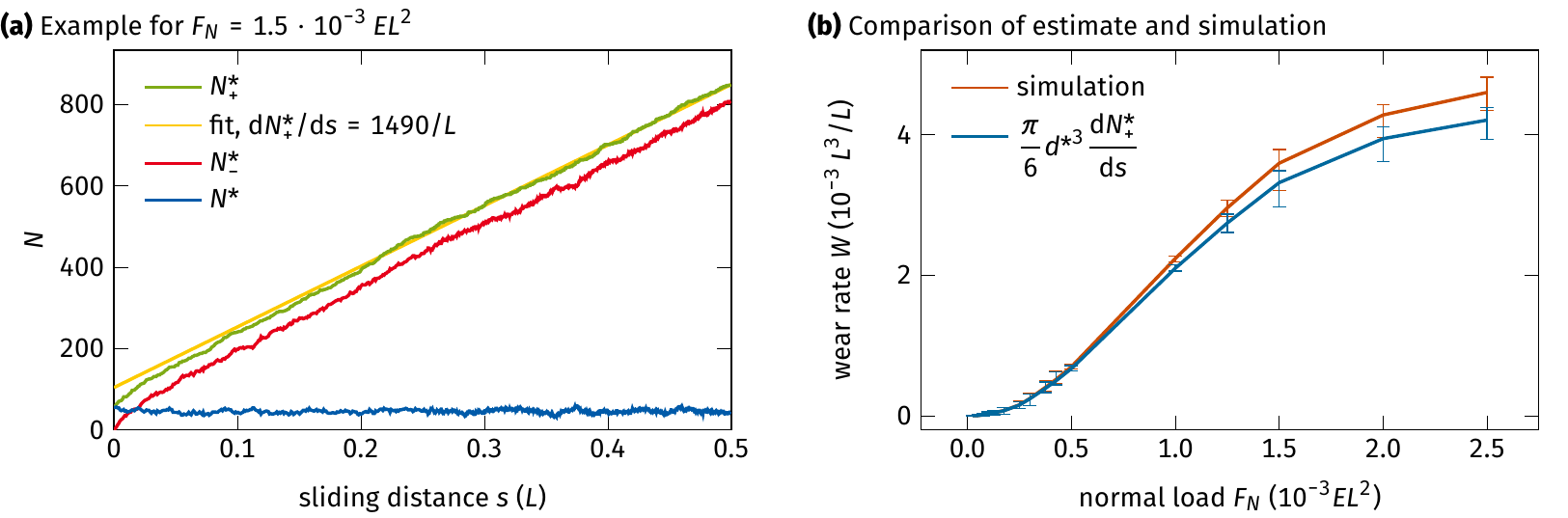}
    \caption{Wear particle formation rate. (a) The rate can be
      obtained by linear regression of the total number of wear
      particles $N^*_+$ formed over a sliding distance $s$. The number
      of contact junctions $N^*$ with $d \geq d^*$ stays roughly
      constant for any $s$. (b) The resulting wear rate compared with
      our simulation results.}
    \label{fig:number-of-wear-particles}
\end{figure}

Equation~\ref{eq:wear-rate-new} indicates the importance of the number
of wear particles $N^*_+$ that are created over a given sliding
distance $s$. Figure~\ref{fig:number-of-wear-particles} presents a
detailed look at this data for one of the main simulations with
$\tilde{\tau} = 0.006\,E$ and $\mathcal{G} = 0.25 \cdot 10^{-6}\,EL$. In
Fig.~\ref{fig:number-of-wear-particles}(a), we see the number of wear
particles formed $N^*_+(s,d^*,F_N)$, the cumulative number of
junctions $N^*_-(s,d^*,F_N)$ that have shrunk below $d^*$ during a
sliding distance $s$, and the number of junctions $N^*(s,d^*,F_N)$
greater than $d^*$ at a given sliding distance $s$. It is
$N^* = N^*_+ - N^*_-$, wherefore $N^*_-$ closely follows $N^*_+$ after
some steps. Just like the wear rate, the wear particle formation rate
is linear with sliding distance. The slope
$\mathrm{d}N^*_+/\mathrm{d}s$ is obtained by linear regression, cf.\
the wear rate in Fig.~\ref{fig:running-in}.
Figure~\ref{fig:number-of-wear-particles}(b) demonstrates that
Eq.~\ref{eq:wear-rate-new} with $\omega = \pi/6$ slightly
underestimates our simulation results obtained by directly summing the
volumes of wear particles, even if the simulation also
assumes spherical particles. The reason is that---due to
discretization---contacts may grow a little bit bigger than $d^*$
before they detach. In reality, particles would not be perfectly
spherical and adjustments to $\omega$ have to be made in any case.

\section{Apparent contact area}
\label{sec:apparent-contact-area}

\begin{figure}[b]
    \centering
    \includegraphics[center]{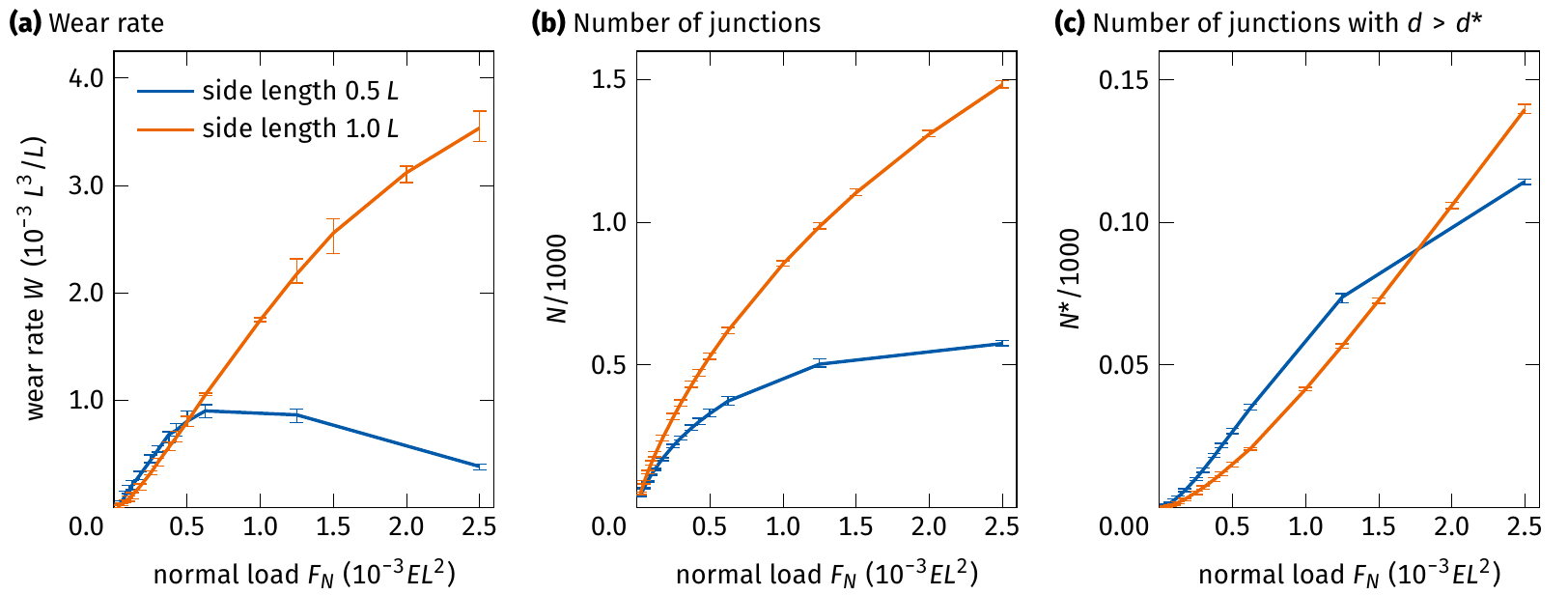}
    \caption{Influence of the apparent contact area. (a) The wear
      rates seem roughly independent of the apparent contact area
      until a certain load, at which the wear rate of the smaller
      surface breaks down. (b) The number of junctions $N$ follows a
      similar trend for both sizes, although the larger surface has
      more individual contacts. (c) The number of junctions $N^*$ that
      are larger than $d^*$, though, starts to collapse at higher
      loads for the smaller surface, explaining the breakdown of the
      wear scaling in (a). The simulations were run 5 times each, the
      error bars indicate the minimum and maximum value of the wear
      rate.}
    \label{fig:apparent-contact-area}
\end{figure}

In order to study the behavior of our model as a
function of the apparent contact area, we also produced five pairs of
surfaces with a side length of $L_1 = 0.5\,L$ (we will not normalize
material properties and results to $L_1$, but stay with $L$ as a
length unit for easier comparison). Otherwise, all surface and
material parameters were the same as for the surfaces with side
length $L$: $H = 0.7$, $\lambda_l = L/256 = L_1/128$,
$\lambda_u = L/16 = L_1/8$, and $h_\text{RMS} = 0.005\,L$, as well as
material properties of $\nu = 0.3$, $\tilde{\tau} = 0.010\,E$, and
$\mathcal{G} = 0.50\cdot10^{-6}\,EL$.

Figure~\ref{fig:apparent-contact-area}(a) shows that the wear rates
are indeed comparable up to normal loads of
$F_N \approx 0.5\cdot10^{-3}\,EL^2$. At higher loads, the larger
surface continues a roughly linear increase of wear rate with load,
while the wear rate of the smaller surface breaks down. Since the wear
rate depends on the number of junctions growing to the critical size,
we also looked at the average number of contact junctions as a
function of load (Fig.~\ref{fig:apparent-contact-area}(b)). This
analysis proves to be not very enlightening on its own, though: The
larger surface has more contact spots, but does not wear more, and
there is no crossover in behavior as a function of normal load. A
comparison of the number of contact junctions $N^*$ that are larger
than the critical size reveals more.  In
Fig.~\ref{fig:apparent-contact-area}(c) we can see that $N^*$ breaks
down from its approximately linear scaling at the same time as the
wear rate does. We interpret this to mean that smaller surfaces
statistically have less contact spots overall and thus cannot
accommodate the necessary number of large contact spots to keep up the
linear growth of wear rate with load. \hyperref[sec:suppl-data]{Video~2}
shows a simulation in the high-load regime of an even smaller surface
and indicates that contact junctions also start to merge, which
additionally reduces the number of individual contacts. This might
well be connected to a transition to severe wear, since the contacts
are expected to interact if they come closer together, at which point
their wear particle sizes are no longer predictable from the
individual contact areas \citep{Aghababaei2018, Pham-Ba2020}. Such
phenomena are not included in the present model, and we have to treat
data after this ``breakdown'' as unreliable.

Note that we chose the larger surface to obtain the results in the
main text since it can sustain a larger range of loads. Because we
originally started our investigations with the smaller surface,
though, it was used for the verification studies in the appendices
unless otherwise noted.

\section{Lower wavelength cutoff of the surface spectrum}
\label{sec:lower-wavelength-cutoff}

\begin{figure}[H]
    \centering
    \includegraphics[center]{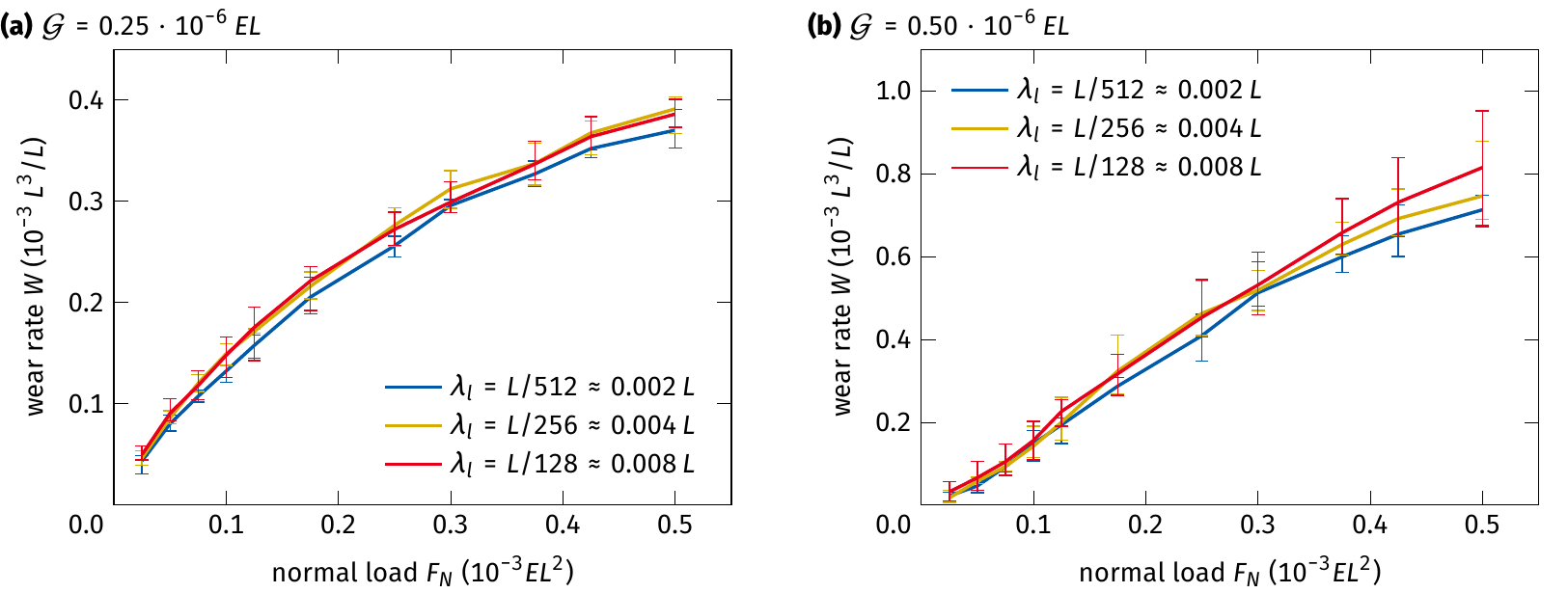}
    \caption{The model is robust against changes of the lower
      wavelength cutoff of the surface spectrum. The simulations were
      run 5 times each, the error bars indicate the minimum and
      maximum value of the wear rate.}
    \label{fig:lower-wavelength-cutoff}
\end{figure}

We studied the influence of the lower wavelength cutoff of the surface
spectrum (Fig.~\ref{fig:lower-wavelength-cutoff}) using surfaces with
roughness parameters of $H = 0.7$, $\lambda_u = L/16$,
$h_\text{RMS} = 0.005\,L$, as well as material properties of
$\nu = 0.3$, $\tilde{\tau} = 0.010\,E$. The parameters $\lambda_l$ and
$\mathcal{G}$ were varied as indicated in the figure. Within the error
of the simulation and the range of values studied, the model is robust
against changes of $\lambda_l$.

\section{RMS of heights}
\label{sec:h-rms}

\begin{figure}[]
    \centering
    \includegraphics[center]{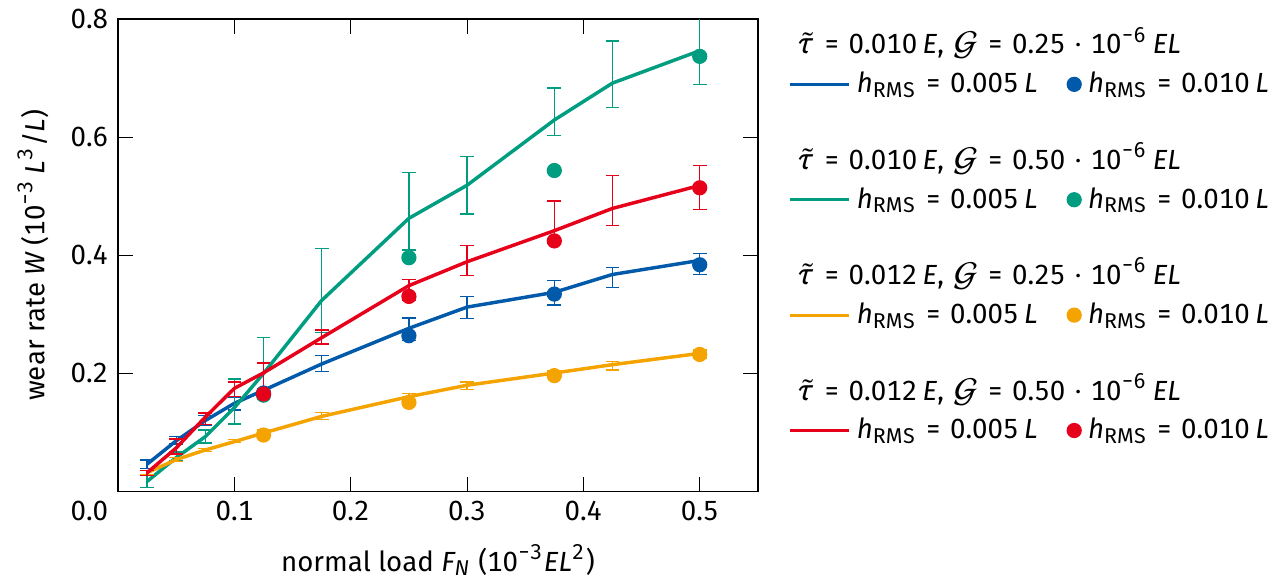}
    \caption{The RMS of heights seems to have at most a very small
      influence. For the increased value of $h_\text{RMS}$, only a
      single surface realization was used, therefore the data is
      presented as points without error bars.}
    \label{fig:h-rms}
\end{figure}

We studied the influence of the RMS of heigths of the surface spectrum
(Fig.~\ref{fig:h-rms}) using surfaces with roughness parameters of
$H = 0.7$, $\lambda_l = L/256$, $\lambda_u = L/16$, as well as
material properties of $\nu = 0.3$ and different $\tilde{\tau}$ and
$\mathcal{G}$ as indicated in the figure. We see that $h_\text{RMS}$
has no influence in the range of values investigated, likely because
the number of junctions matters more and because high peaks are
flattened by plasticity.

\section{Wear law based on ``contact severity''}
\label{sec:Sc}

\begin{figure}[H]
  \centering
  \fcapside[9cm]{%
    \caption{Relation between our simulated wear rates and the
      proposed ``contact severity'' parameter $S_c$
      \citep{Kim1986}. The dashed line represents the best fit of all
      data.}
    \label{fig:Sc}%
  }{%
    \includegraphics[center]{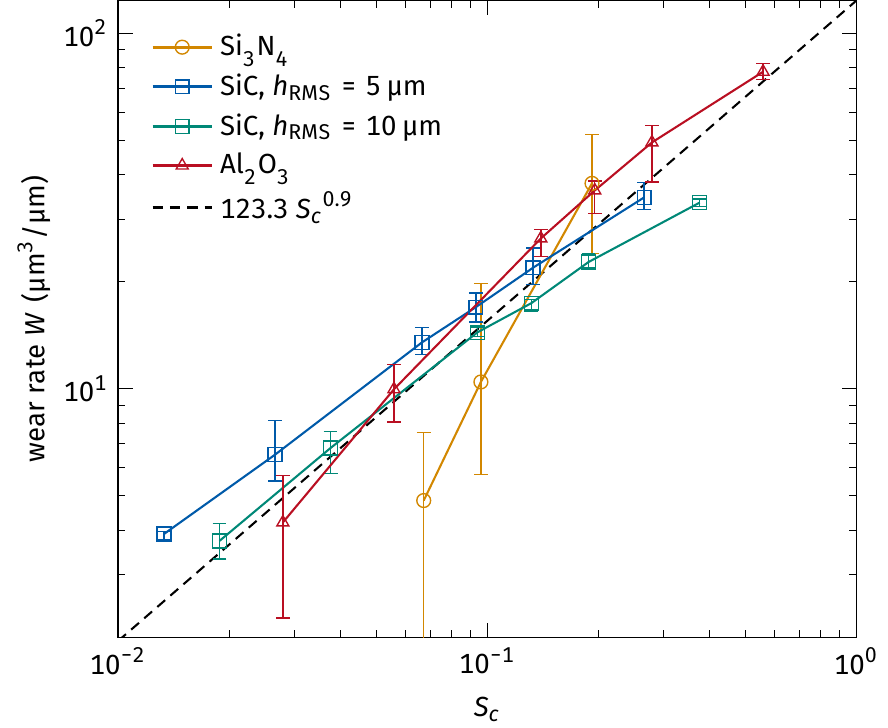}%
  }
\end{figure}

For the sake of completeness, we will shortly comment on a proposed
wear parameter $S_c$ in the publication of \citet{Kim1986}, where
\begin{equation}
  \label{eq:Sc-kim-fit}
  W = \alpha {S_c}^n
\end{equation}
with
\begin{equation}
  \label{eq:Sc-kim}
  S_c = p_N \dfrac{\sqrt{h_\text{RMS}}}{K_\text{Ic}}.
\end{equation}
In contrast to \citet{Archard1953}, the hardness does not play a role,
but the toughness does. The experimental data in \citet{Kim1986}
consists of only two loads per material and Eq.~\ref{eq:Sc-kim-fit} is
fitted to a log--log plot, where it works approximately, with values
of $\alpha = \SI{1.56e-5}{mm^3/mm}$ and $n = 5.46$. The same is true
for our simulations, see Fig.~\ref{fig:Sc}, although with fit
parameters of $\alpha = \SI{1.233e-4}{mm^3/mm}$ and $n = 0.9$ and some
deviation for Si$_3$N$_4$. It is not clear if Eqs.~\ref{eq:Sc-kim-fit} and
\ref{eq:Sc-kim} are predictive, especially since we do not find a
strong dependence on $h_\text{RMS}$, which seems to have been included
for the convenience of having a unitless $S_c$. The influence of
hardness is completely ignored. Indeed, later publications did not
find a clear correlation between $S_c$ and $W$ and rather propose that
the maximum stress $p_N$ leads to the correlation, similar to $F_N$ in
Archard's wear law \citep{Hsu2004}. It is therefore safe to say that
such a model does not fare better than other empirical wear laws
and an agreement between our values for $\alpha$ and $n$ and the ones
of \citet{Kim1986} should not be expected.

\section{Supplementary data}
\label{sec:suppl-data}

Supplementary data related to this article can be found at the end of
this document. Videos are freely available at
\href{https://arxiv.org/abs/2004.00559}{https:/\!/arxiv.org/abs/2004.00559}
and \href{https://doi.org/10.1016/j.jmps.2020.104238}
{https:/\!/doi.org/10.1016/j.jmps.2020.104238}.

\clearpage
\includepdf[pages={1,2,3}]{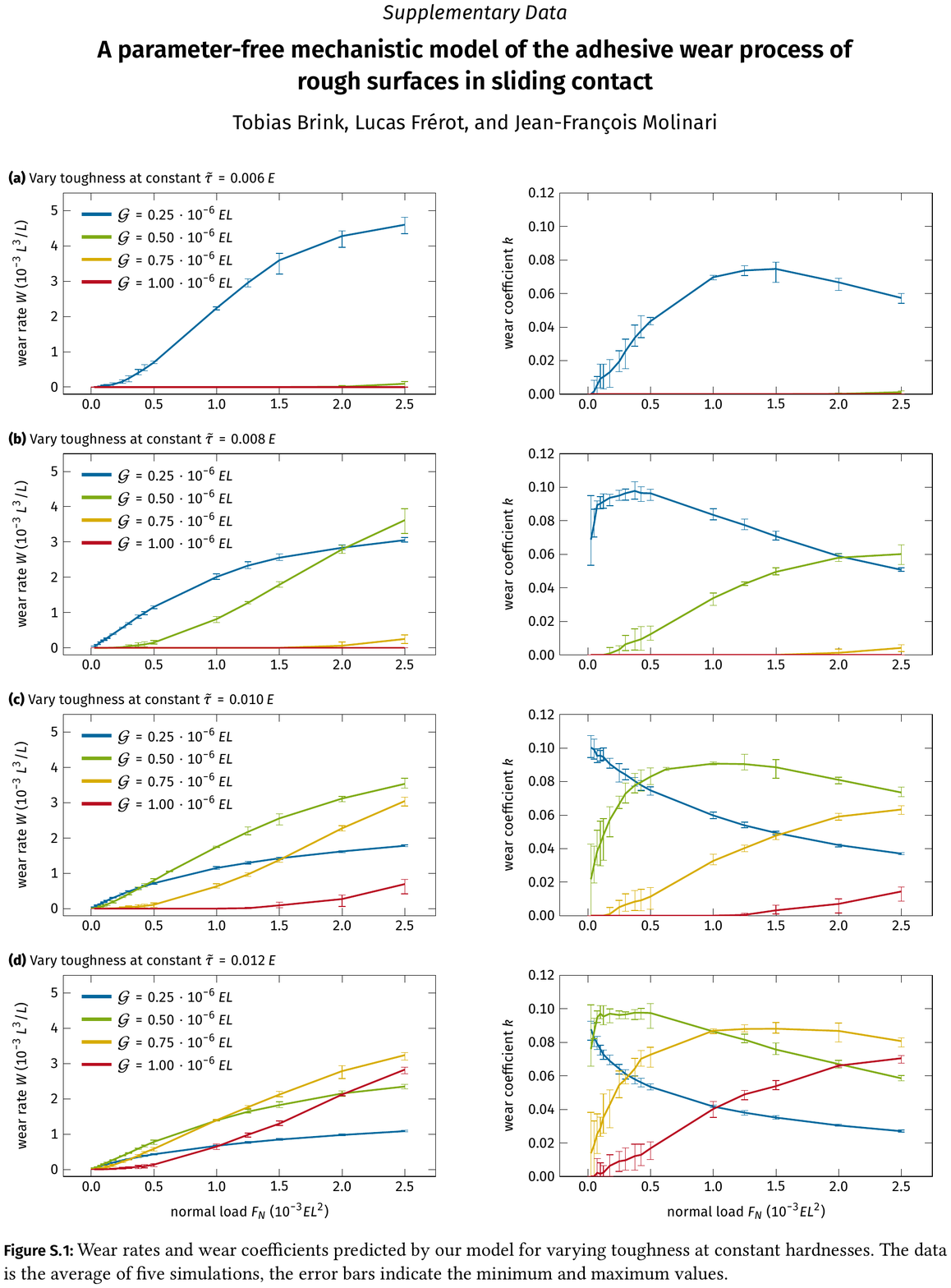}

\end{document}